# Room temperature operation of germanium-silicon single-photon avalanche diode

Neil Na*, Yen-Cheng Lu, Yu-Hsuan Liu, Po-Wei Chen, Ying-Chen Lai, You-Ru Lin, Chung-Chih Lin, Tim Shia, Chih-Hao Cheng, and Shu-Lu Chen

[1]Artilux Inc., Zhubei City, Hsinchu County, Taiwan, ROC
*Email: neil@artiluxtech.com

**The capability of detecting single photons has greatly advanced a wide range of research fields[1-11]. While various types of single-photon detectors have been developed[12], due to two main factors, i.e., 1) the need of operating at cryogenic temperature[13,14], and 2) the incompatibility with complementary metal-oxide-semiconductor (CMOS) fabrication process[15,16], so far only Si-based single-photon avalanche diode (SPAD)[17,18] has gained mainstream success and been used in consumer electronics. With growing demand in migrating the operation wavelength from near-infrared (NIR) to short-wavelength infrared (SWIR) for better safety and performance[19-21], alternative solution is required since Si has negligible optical absorption for wavelengths beyond 1 μm. Here we report a CMOS compatible, high-performing GeSi SPAD operated at room temperature, featuring a noise-equivalent power (NEP) improvement over the previous Ge-based SPADs by 2 to 3.5 orders of magnitude. Key parameters such as dark count rate (DCR), single-photon detection probability (SPDP) at 1310 nm, timing jitter, after-pulsing characteristic time, and after-pulsing probability (APP), are respectively measured to be 19 kHz/μm$^2$, 12 %, < 188 ps, 90 ns, and < 1%, with a low breakdown voltage of 10.3 V and a small excess bias of 0.75 V. Three-dimensional (3D) point-cloud (PCL) images are captured with direct time-of-flight (TOF) technique as proof of concept. This work paves the path toward deploying single-photon sensitive, SWIR sensors, imagers, and photonic integrated circuits for daily life applications.**

Detectors that are extremely sensitive and may discern the smallest quanta of light, i.e., a single photon, are referred as single-photon detectors. These detectors have played crucial roles during the past decades in scientific researches on optical quantum computation[1,2], quantum key distribution[3,4], quantum imaging[5,6], and bio photonics[7], by detecting the granular properties of light, and in recent technological developments of LiDAR in coaxial setting with single detectors[8,9], or in camera setting with pixel arrays[10,11], by time-resolving the photon arrival events. Solid-state implementations of such detectors through superconductors, e.g., superconducting nanowire single-photon detector (SNSPD), and semiconductors, e.g., SPAD or sometimes referred as Geiger-mode avalanche photodiode (APD), are perhaps the most well-known examples. While SNSPD may detect a broad wavelength range from mid-infrared to visible, it typically operates at a cryogenic temperature of a few K[12-14] that limits its usage mainly as a lab instrument. On the other hand, SPADs based on InGaAs or Si may only detect specific wavelength range from SWIR to NIR or from NIR to visible, respectively, but they can be operated at near room temperature or room temperature[12,15-18] and hence become popular for general purposes. In particular, Si-based SPAD is intrinsically CMOS compatible, where homogenous integration with CMOS circuits (i.e., monolithic) at sub-micron technology node under front-side illumination[17], as well as heterogenous integration with CMOS circuits (i.e., wafer-to-wafer bonding) at deep sub-micron technology node under back-side illumination[18], have been demonstrated. Such powerful advancements have made Si-based SPAD enter the segment of mainstream consumer electronics as proximity sensors and 3D imagers at NIR. Recently, SWIR has emerged as the next-generation choice of wavelength due to the following reasons: 1) better laser eye-safety since NIR can cause damage to retina[19], 2) lower ambient light interference since solar spectrum is weaker at SWIR or even absent around 1380 nm wavelength[20], 3) higher atmospheric transmission since free-space optical transmissivity is dominated by Rayleigh scattering[21], and 4) less interference with Si devices and circuits as Si absorbs NIR but not SWIR. Consequently, SWIR single-photon detectors operated at room temperature and being CMOS compatible are actively sought after. In the literature, Ge-based SPAD may be a potential candidate, but it either experimentally shows high DCR and/or low SPDP so that decent performance can only be reached at a low temperature typically < 200 K, or is studied only through theoretical simulations without experimental supports.

## SPAD device benchmark and concept

In this paper, we report the fabrication of a 15 μm diameter GeSi SPAD with CMOS compatible fabrication process, and characterize it at room temperature to demonstrate its high performance. To benchmark the demonstrated GeSi SPAD with the Ge-based and InGaAs-based SPADs in the literature, we calculate the NEP by using the definition[31]

$$NEP = \frac{\hbar\omega\sqrt{2 \cdot DCR}}{SPDP}, \quad (1)$$

where the DCR and SPDP from the highest operation voltages at the highest operation temperatures are used in the calculations. In Fig. 1, the operation temperature and NEP of Ref. [22-28] (Ge-based), Ref. [15,16] (InGaAs-based), and this work are plotted. To better compare the NEP spread over wide ranges of temperatures, without loss of generality, DCR is assumed to double for every 10 K increase in temperature[31], which is typical for Ge-based and InGaAs-based devices. Then, the dashed-dotted arrows in Fig. 1 are added to indicate the NEP of the previously reported Ge-based and InGaAs-based SPADs if they were operated at 300 K: the demonstrated GeSi SPAD features an unprecedentedly low NEP around a few fW/√Hz at room temperature, which is about 2 to 3.5 orders of magnitude lower than those of the previously reported Ge-based SPADs. The drastic difference is mainly contributed to the low dark current of the demonstrated GeSi SPAD. Taking Ref. [24] as an example, which is the first demonstration of Ge-on-Si SACM APD for SPAD, it features a dark current about 1 μA at the punch-through point around -22 V, whereas the demonstrated GeSi SPAD features a dark current about 20 pA at the punch-through point around -9 V. Note that in Fig. 1 the device areas are not normalized, because for the previously reported Ge-based SPADs the DCR depends highly on design and process and does not scale well with the device area (e.g., Ref. [27] has an area about 15 times larger than that of Ref. [28] but their NEP are about the same). Still the demonstrated GeSi SPAD has a NEP that is on average 2.5 orders of magnitude lower than those of the previously reported Ge-based SPADs if the device areas are normalized. On the other hand, the NEP of the demonstrated GeSi SPAD is catching up the NEP of the previously reported InGaAs-based SPADs with a difference only about 0.5 orders of magnitude. Moreover, the after-pulsing of the demonstrated GeSi SPAD is much minor than those of the previously reported InGaAs-based SPADs, suggesting for certain applications the demonstrated GeSi SPAD may already outperform some of the previously reported InGaAs-based SPADs (will be elaborated in the later paragraph on after-pulsing).

The significant reduction of the dark current of the demonstrated GeSi SPAD is made possible by addressing the following issues: 1) the threading dislocations occurring on the (311) planes in Ge arisen from the 4 % lattice-mismatch between Ge and Si, 2) the lack of high-quality surface passivation agents over Ge surfaces, and 3) the high electric fields present in the low-quality material regions especially along the Ge sidewalls. For the issue of threading dislocation, various methods have been pursued such as graded buffer layers grown between Si and Ge[32], post-Ge high-temperature thermal annealing[33], low-energy plasma-enhanced chemical vapor deposition (CVD)[34], high-aspect-ratio trapping of defects[35], and Ge-on-insulator with defects chemically-mechanically polished[36], to name a few. For the issue of surface passivation, various agents have been experimented such as amorphous Si[37], $GeO_2$[38], $Al_2O_3$[36], and pure Ga/B[39], to name a few. Generally, with the combined efforts of reducing threading dislocations and surface dangling bonds, Ge-on-Si photodiode (PD) featuring a low dark current density down to hundreds or even tens of $\mu A/cm^2$ can be accomplished at a low reverse-bias voltage. However, to be used as a high-performing APD or even SPAD, such a low dark current must be maintained while the device is biased above the unity gain point, in which the high electric fields needed for transporting photo-carriers and generating impact ionizations kick in. In fact, the previously reported Ge-based APDs[40-45] did not meet such a criterion, and here the demonstrated GeSi SPAD achieves this through a novel design to effectively screen the high electric fields off the low-quality Ge sidewalls. The concept is illustrated in Fig. 2 (a), showing the schematic plot of the demonstrated GeSi SPAD. Fig. 2 (b) shows the cross-the corresponding scanning-electron-microscope (SEM) image. The device consists of elements such as p-contact doping region (deepest blue) in Ge, p-type charge layer (lighter blue) in Si, and deep n-well doping region connected to n-contact doping region through n-sinker doping region all in Si. Then, a p-type screening layer (deeper blue) in Si is implemented at regions above the charge layer and below the Ge sidewalls (for more fabrication details see Methods). The dopant density of the screening layer must be high enough so that when the charge layer is depleted, the high electric fields below the screening layer would terminate at the screening layer, i.e., would not penetrate the Ge sidewalls. At the same time, the dopant density of the screening layer must be low enough so that pre-mature breakdown below the screening layer would not occur, i.e., the breakdown electric fields below the charge layer are uniform regardless if the screening layer above presents or not. Indeed, through electrostatic simulation, we observe negligible electrical fields < 1 kV/cm along the Ge sidewalls even when the device is biased far above its breakdown voltage.

### SPAD device characterization

Fig. 2 (c) shows the IV measurement from wafer-level testing where the reverse-bias current is plotted as a function of the applied voltage, with and without a collimated 1310 nm laser beam to illuminate the device under testing (DUT). Two DUT samples named S1 and S2 are given, and they are of the same 15 μm diameter device except the charge layer concentration of S1 is slightly larger than that of S2. A reference PD, i.e., the same 15 μm diameter device except its charge layer is removed, shows a photocurrent about 1 μA under the same measurement conditions. By using this number and intercepting the photocurrents in Fig. 2 (c), the corresponding primary (i.e., at unity gain point) dark currents of S1 and S2 are determined to be about the same 8 pA at -9.2 V and -8.8V, respectively, corresponding to 4.5 μA/cm$^2$ and is the lowest value ever reported for Ge-based APD and/or SPAD. Additional IV measurements show that the primary dark current is increased by about 12 times when the device diameter is increased from 15 μm to 50 μm, suggesting that the primary dark current scales reasonably well with the device area. The punch-through voltage $V_{PT}$ of S1 and S2 are determined to be -9.3 V and -9 V, respectively, from additional CV measurements and coinciding with the "knee" points of the photocurrents in Fig. 2 (c). The breakdown voltage $V_{BD}$ of S1 and S2 are determined to be -10.3 V and -10.9 V, respectively, by defining 10 μA dark current as the breakdown point. Moreover, the breakdown voltage thermal coefficient, defined as $(\Delta V_{BD}/V_{BD})/\Delta T$ where $\Delta V_{BD}$ and $\Delta T$ are the increments of breakdown voltage and temperature, is determined from additional temperature-dependent IV measurements to be about 0.06 %/°C between 30°C and 80°C, which agrees reasonably with the case of Si pin APD[46]. The voltage differences between $V_{BD}$ and $V_{PT}$ for S1 and S2 are 1 V and 1.9 V, respectively, which manifest the difference in charge layer concentrations. Interestingly, it can be observed that the shapes of the photo IV curve and the corresponding dark IV curve in Fig. 2 (c) share striking similarities, so we plot the photocurrent gain and the corresponding dark current gain as a function of the applied voltage in Fig. 2 (d) for comparison. It is found that they overlap closely with each other up to gain about a hundred, suggesting that the multiplied component of the dark current originates at places away from the low-quality Ge sidewalls due to the screening layer.

To characterize the performance of the demonstrated GeSi SPAD, gated-mode operation[22-28] is applied in which the cathode of the DUT is biased repeatedly at repetition time $T_r$ through a bias-Tee with a direct-current (DC) voltage below $V_{BD}$ and a radio-frequency (RF) voltage pulse causing a total operation voltage above $V_{BD}$ during gate time $T_g$ (for more measurement details see Methods). The count rate directly reported by the counter or the oscilloscope should be multiplied by $T_r/T_g$ to obtain the actual DCR. Since $T_g$ is typically around a few or tens of nanoseconds, DCR below GHz or sub-GHz range can be time-resolved and hence gated-mode operation has been used in characterizing SPADs that may have wide ranges of DCRs. In Fig. 3 (a), the measured DCR from die-level testing using the counter is plotted as a function of excess bias $V_{ex}$ (i.e., the difference voltage between the operation voltage above $V_{BD}$ and $V_{BD}$) for S1 and S2, given $T_g$ and $T_r$ equal to 5 ns and 102.4 ns, respectively. DCR between 1.5 MHz and 14.1 MHz (i.e., about 8.5 kHz/μm$^2$ and 79.8 kHz/μm$^2$) can be reached for $V_{ex}$ between 0.5 V and 2.5 V (i.e., about 2 % and 23 % of $V_{BD}$ ), which are the lowest values ever reported for Ge-based SPAD. Next, 1310 nm picosecond laser pulses in synchronization with gate pulses are injected into a confocal microscope setup that couples to the device (for more measurement details see Methods). The average number of photons per pulse $\langle n \rangle$ received by the DUT is controlled by first calibrating the optical loss of coupling the confocal microscope to the device and then adjusting a variable optical attenuator. SPDP can be obtained by subtracting the measured DCR from the measured total count rate (TCR), i.e., the count rate with the pulsed laser turned on, and multiplying it by $T_g$/<n> assuming <n><<1 to ensure the probability for a laser pulse containing multiple photons is negligible. In Fig. 3 (b), the measured SPDP from die-level testing using the counter is plotted as a function $V_{ex}$ for S1 and S2, given the same conditions of $T_g$ and $T_r$ as in Fig. 3 (a). SPDP between 5 % and 24 % can be reached for $V_{ex}$ between 0.5 V and 2.5 V given <n>=0.1. In Fig. 3 (c), the measured timing jitter $\delta$ in full-width-half-maximum (FWHM) from die-level testing using the oscilloscope is plotted as a function $V_{ex}$ for S1 and S2, given the same conditions of $T_g$ and $T_r$ as in Fig. 3 (b). $\delta$ between 115 ps and 207 ps can be reached for $V_{ex}$ between 0.5 V and 2.5 V given <n>=10. Note the condition <n>=10 is chosen for satisfying a signal-to-background noise ratio equal to 10 (assuming SPDP ~ 10 %, DCR ~ 2 MHz, $T_g$ ~ 5 ns) commonly required for single-shot LiDAR. Exemplary voltage waveforms from the DCR measurement of S1 in Fig. 3 (a) operated at $V_{ex}$ equal to 1 V, 1.5 V, and 2 V are shown in Fig. 3 (d), where sharp signal rise due to breakdown and sharp signal fall due to gate-on turned gate-off are observed. Exemplary photon arrival histogram from the $\delta$ measurement of S1 in Fig. 3 (c) operated at $V_{ex}$ equal to 2 V is shown in Fig. 3 (e), where $\delta$ of 115 ps is observed and attributed mainly to the device $\delta$ since the oscilloscope system $\delta$ is estimated to be around a few tens of picoseconds and is negligible. Note the shape of the histogram is exponentially-decreasing with time because the device $\delta$ is dominated by the device long RC time constant (roughly between 100 and 200 ps) due to higher contact resistance and parasitic capacitance. Finally, as shown in Fig. 3 (f), DCR and SPDP of S1 and S2 are measured from wafer-level testing using the counter at $V_{ex}$ equal to 1 V, 1.25 V, 1.5 V, 1.75, and 2 V, when the chuck is controlled to 20 °C, 30 °C, and 40 °C temperatures. For DCR, its increase with increasing temperature is about 2.5 times for every 10 °C, corresponding to an activation energy about 0.64 eV and is slightly smaller than the Ge bandgap about 0.66 eV. This suggests that defects in Ge with energy levels close to Ge band edges are responsible for generating DCR, and so there is still room to further minimize DCR through optimizing the fabrication process. As for SPDP, it stays roughly the same throughout the controlled temperature range.

When a breakdown event occurs and induces strong current flow passing through the multiplication region, some carriers may be trapped in the defects and subsequently released at a time delay causing another breakdown event. Such a phenomenon is called after-pulsing, and its characteristic time $\tau$ and occurrence probability APP are crucial parameters in SPAD operations. Here we characterize the after-pulsing of the demonstrated GeSi SPAD at room temperature, by applying time-correlated counting under gated-mode operation to measure the statistics of interarrival time with dark counts[47,48] (for more measurement and calculation details see Methods). In Fig. 3 (g), the measured histograms for interarrival time normalized by $T_r$ are shown for several $V_{ex}$, given the same conditions of $T_g$ and $T_r$ as in Fig. 3 (a)-(f). The black dashed lines are fitting curves for the thermally-generated counts $n_{TG}(i)$, whereas the after-pulsing counts $n_{AP}(i)$ are above the fitting curves and distribute exponentially. $i$ is the index representing the $i$-th gate time after the reference gate time. In Fig. 3 (h), the calculated $\tau$ is shown as a function of $V_{ex}$. It is observed that $\tau$ is relatively insensitive to $V_{ex}$ and is about 90 ns for $V_{ex}$ between 1 V and 2 V. In Fig. 3 (i), the calculated $APP(1)$ (i.e., $i$=1), $APP(2)$ (i.e., $i$=2), and $APP^t$ (i.e., the total APP) are shown as a function of $V_{ex}$. It is observed that $APP^t$ is exponentially related to $V_{ex}$ and is between 0.7 % and 1.9 % for $V_{ex}$ between 1 V and 2 V. The numbers of $\tau$ and APP are found to be in reasonable agreement with those of the Si-based SPADs at room temperature[49], suggesting the high crystalline quality of the demonstrated GeSi SPAD. Note for certain applications such as multi-shot LiDAR based on direct TOF technique using pulse rather than continuous wave (CW), in order to reach a target signal-to-background noise ratio, the integration time required is proportional to $NEP\sqrt{\tau}$. Consequently, the performance of the demonstrated GeSi SPAD may already outperform some of the previously reported InGaAs-based SPADs (considering *NEP* about 3.5 times lower from Fig. 1; $\sqrt{\tau}$ about 3 to 33 times higher from Ref. [15,16]) in this and other similar cases.

## LiDAR system demonstration

As proof of concept, the setup for producing data shown in Fig. 3 is modified to apply the demonstrated GeSi SPAD for LiDAR based on direct TOF technique. 3D scanning and imaging are done by replacing the variable optical attenuator with a coaxial transceiver that uses x-y galvo-mirrors to steer the fired laser pulses and collect the returned laser pulses along the same optical path (for more measurement details see Methods). As shown by the 2D RGB image in Fig. 4 (a), the 3D objects are placed against the wall at different distances away from the mirrors under regular indoor ceiling lighting. Alphabet letters standing on the lower cardboard staircase are of 100 mm in height, and the platonic solids standing on the upper cardboard staircase are of 65 mm in height. In Fig. 4 (b), the captured 3D PCL image by using sample S1 at $V_{ex}$ equal to 2 V is shown. The relative distances away from the mirrors are color-coded for better visualization, and it can be clearly seen that the spatial orientations of the 3D objects are nicely revolved without any post-processing, attesting to the potential of the demonstrated GeSi SPAD for LiDAR based on direct TOF technique. Two TOF histograms recorded for paths P1 and P2 are shown in Fig. 4 (c), corresponding to the paths that scatter the laser pulses at locations indicated by the 2 blue circles and the 1 red circle marked in Fig. 4 (b). $\langle n \rangle$ received by the GeSi SPAD are estimated to be 0.02 and 1.5 for P1 and P2, respectively. $\delta$ of the 2 peaks for P1 and the 1 peak for P2, after removing the background dark counts and counter system $\delta$ around 99 ps, are found to be about the same 294 ps, which is larger than the $\delta$ shown in Fig. 3 (e) with $\langle n \rangle=10$ by a factor of 2.56.

## Summary and outlook

We have demonstrated the Geiger-mode operation of a high-performing GeSi APD as SPAD at room temperature, which was previously deemed not possible and shall be a strong contender for room temperature operated, CMOS compatible, SWIR single-photon detectors. Since Ge-based detector operated in linear-mode has been commercialized through Si photonic platform[37] and CMOS image sensor plaform[50], the capability of operating in Geiger-mode shall boost the developments of optoelectronic devices and circuits over these platforms toward unprecedented single-photon level sensitivity. Future directions include 1) scaling the device area for developing a sensing/imaging pixel with a pitch smaller than 10 µm, 2) minimizing the contact resistance and the parasitic capacitance, and 3) heterogenous integration with CMOS circuits through wafer-level bonding.

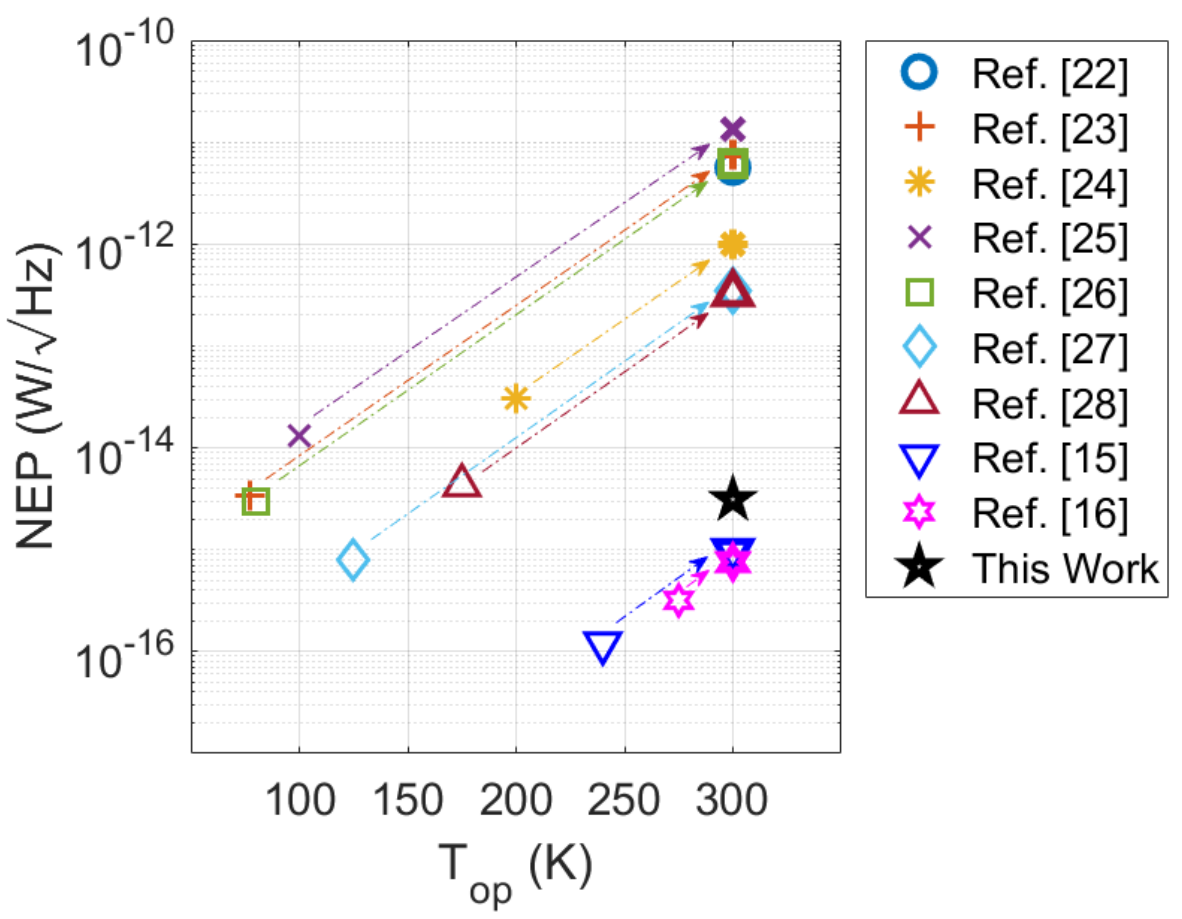

**Fig. 1 | NEP and operation temperature of the previously reported Ge-based (Ref. [22-28]) and InGaAs-based (Ref. [15-16]) SPADs.** The dashed-dotted arrows indicate the NEP of the previously reported SPADs if they were operated at 300 K, assuming DCR is doubled for every 10 K increase in temperature. Note that all devices are tested at 1310 nm wavelength except for Ref. [22] at 1210 nm. The device diameters, following the order of the references shown in the legend, are 120 µm, 40 µm, 30 µm, 25 µm, 4.5 µm (effective), 100 µm, 26 µm, 25 µm, 10 µm, and 15 µm, respectively.

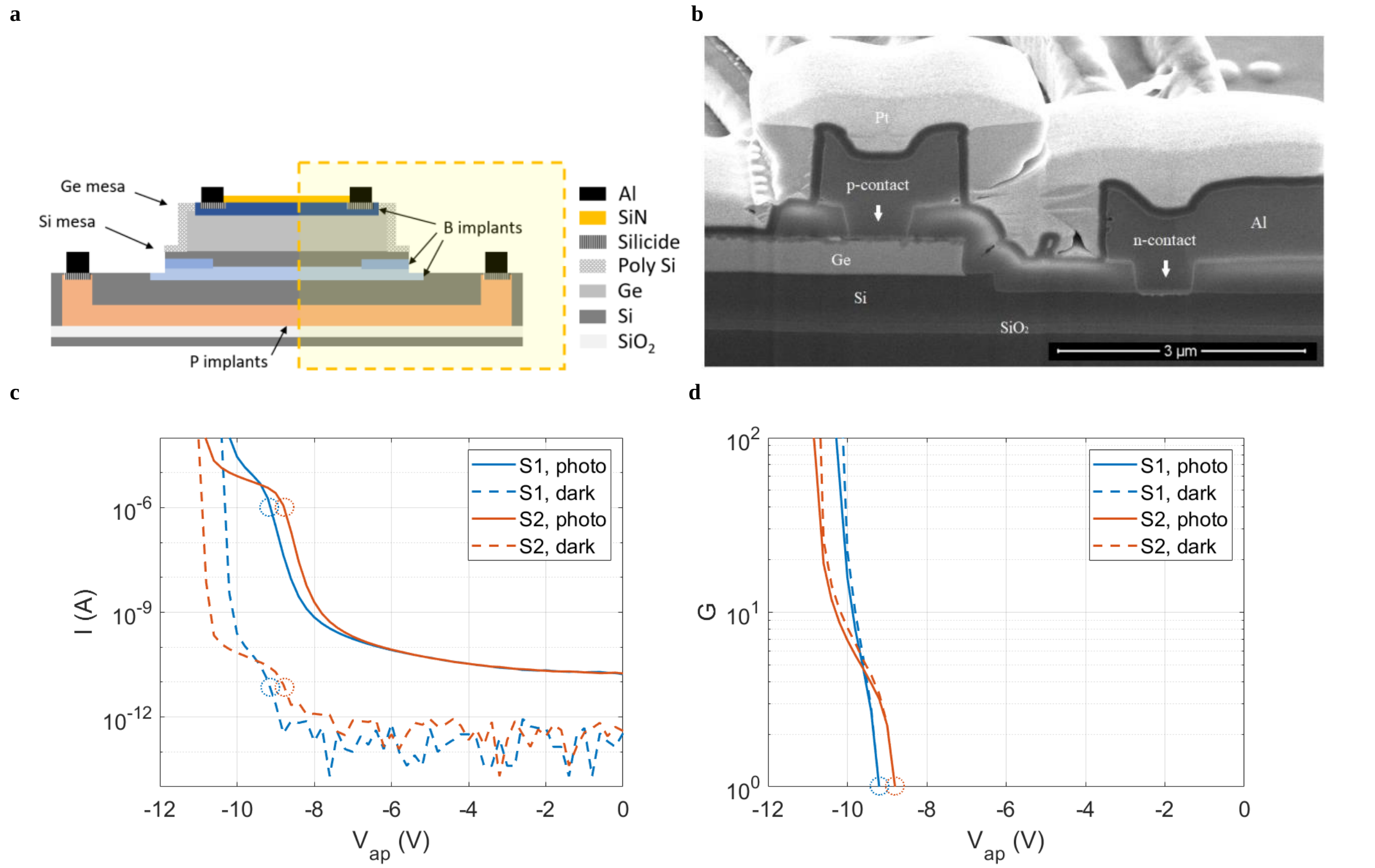

**Fig. 2 | Cross-sectional schematic plot, SEM image, and I-V characteristics of the demonstrated GeSi SPAD. a**, Schematic plot. B implants are used to create the p-contact region (deepest blue) in Ge, and P implants are used to create the n-contact region in Si. Additional B implants in Si form two regions below Ge, one as the charge layer (lighter blue) and one as the screening layer (deeper blue). Additional P implants in Si form the deep n-well region above the buried $SiO_2$, and the n-sinker region between the n-contact region and the deep n-well region. **b**, SEM image. The image is captured corresponding the yellow area enclosed by the yellow dashed lines in Fig. 2 (a). Note that, Pt, i.e., the white conformal layer, is deposited over the device to avoid SEM charging. **c**, **d**, Photo currents (solid curves) and dark currents (dashed curves) as well as their corresponding gains, plotted as a function of applied voltage, for samples S1 (blue) and S2 (red). Determined by the photo current of a reference PD, the primary photo/dark currents in Fig. 2 (c) and the unity gain points in Fig. 2 (d) for samples S1 and S2 are respectively identified and labeled by the blue and red dotted circles.

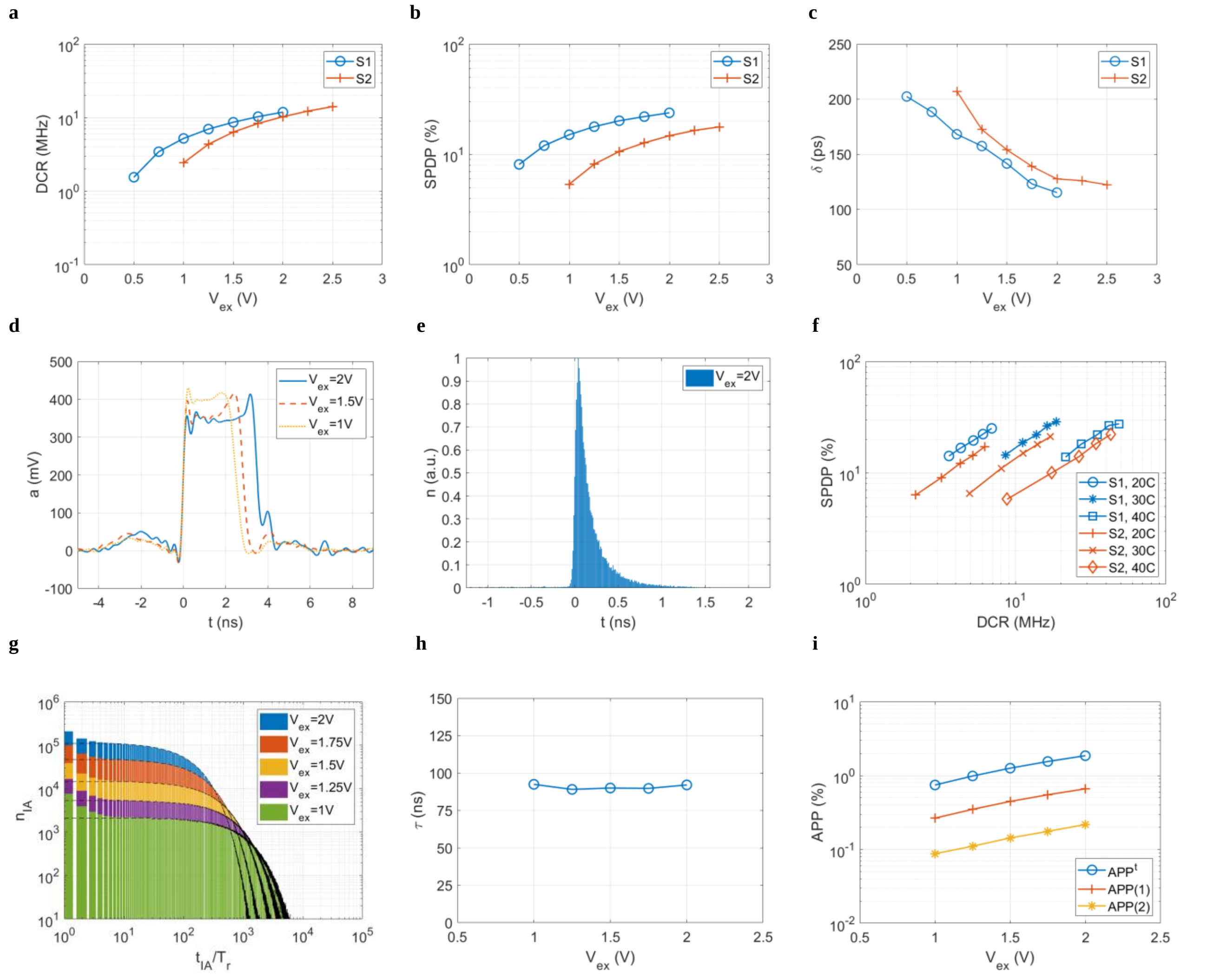

**Fig. 3 | Counting properties of the demonstrated GeSi SPAD.** **a**, Dark count rate DCR measured as a function of $V_{ex}$ for samples S1 and S2. **b**, Single-photon detection probability SPDP measured as a function of $V_{ex}$ for samples S1 and S2, given $<n>=0.1$. **c**, Timing jitter $\delta$ in FWHM measured as a function of $V_{ex}$ for samples S1 and S2, given $<n>=10$. **d**, Exemplary voltage waveforms from the DCR measurement of sample S1 plotted as a function of time. **e**, Exemplary photon arrival histogram from the $\delta$ measurement of sample S1 plotted as a function of time. **f**, Temperature-dependent DCR and SPDP measured as a function of $V_{ex}$ equal to 1V, 1.25V, 1.5V, 1,75V, and 2V for samples S1 and S2, given 20 °C, 30 °C, and 40 °C chuck temperatures. **g**, Interarrival count $n_{IA}$ plotted as a function of interarrival time $t_{IA}$ normalized by $T_r$. **h**, After-pulsing characteristic time $\tau$ calculated with equations in Methods, and plotted as a function of $V_{ex}$. **i**, After-pulsing probability APP calculated from Eq. (3) with equations in Methods, and plotted as a function of $V_{ex}$. In all cases $T_g$ and $T_r$ are set to be 5 ns and 102.4 ns, respectively.

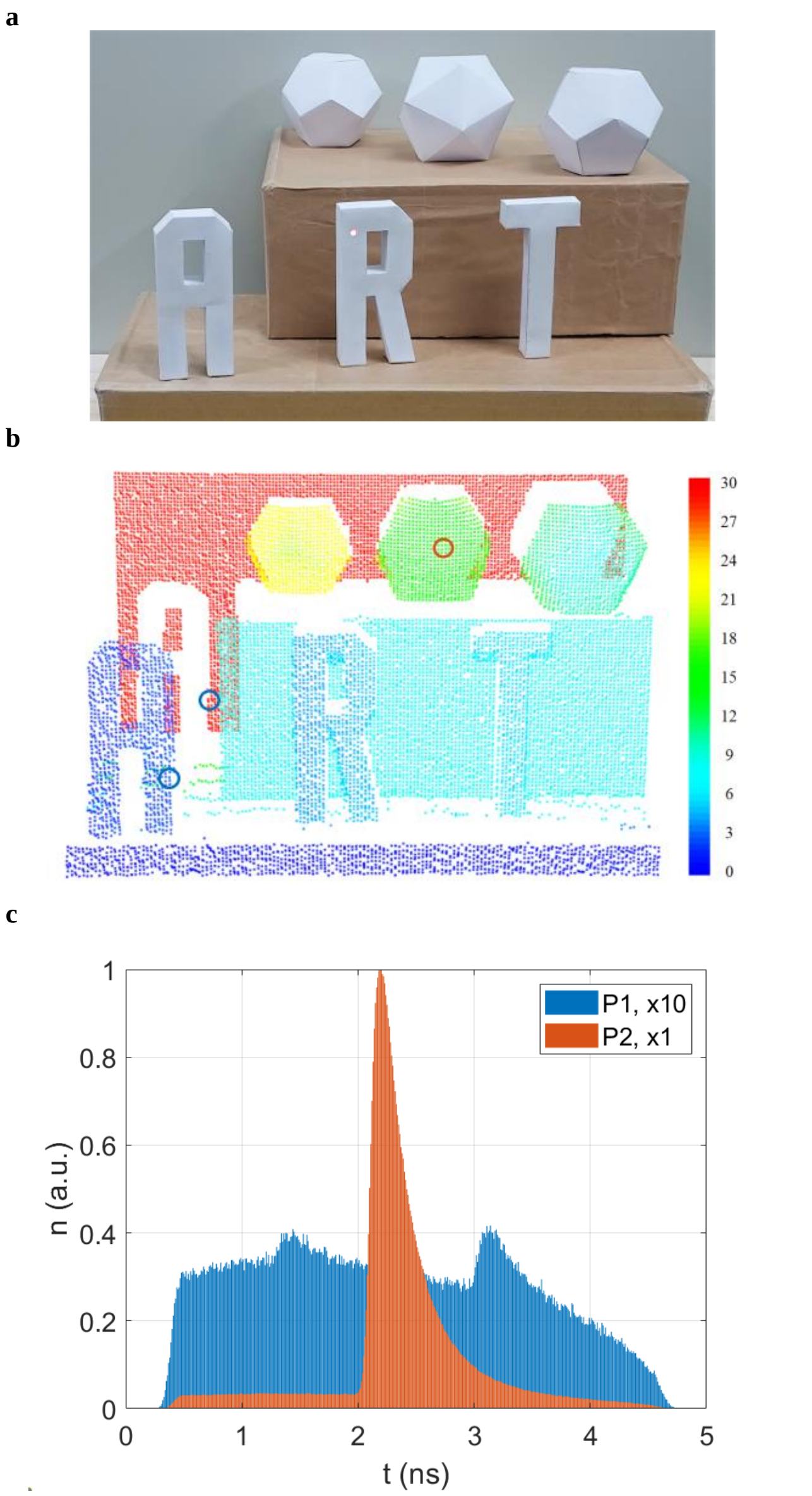

**Fig. 4 | Experimental demonstration of 3D scanning and imaging. a**, The RGB image of the scene. The red laser spot on the alphabet letter R marks the initial point of steering galvo-mirrors. **b**, The PCL image of the scene, captured by the demonstrated GeSi SPAD with direct TOF technique. The color bar indicates the relative distance away from the mirrors in centimeter. **c**, The normalized TOF histograms for path 1 (P1) and path 2 (P2), where P1 and P2 correspond to the paths that scatter the 1310 nm laser pulses at locations marked in Fig. 4 (b) with 2 blue circles and 1 orange circle, respectively. The normalized TOF histogram for P1 is scaled 10 times so that P1 and P2 can be visually compared more easily.

## Methods

### Fabrication

The device fabrication starts with phosphorus (P) implants over a silicon-on-insulator (SOI) wafer to form the deep n-well region and the lower part of the n-sinker region in Si, followed by an epitaxial growth of 250 nm Si (multiplication layer) on the SOI wafer with ultrahigh-vacuum CVD. The buried $SiO_2$ of the SOI wafer is used to isolate the leakage current from the bulk Si substrate. Boron (B) implants form the charge layer and the screening layer around 2.5 μm in width in Si, and P implants form the upper part of the n-sinker and the n-contact regions in Si, followed by another epitaxial growth of buffer Si and 450 nm Ge (absorption layer) between 300 °C and 600 °C. After a post-epitaxy thermal annealing between 600 °C and 800 °C, a series of dry and wet etches form a Ge mesa and a Si mesa, where polycrystalline-Si based passivation is processed over the Ge mesa. B implants form the p-contact region in Ge, followed by silicide process to create the ohmic contacts over both n-contact and p-contact regions, and Al metallization process to define the electrical traces and pads for electrical probing. Finally, SiN is deposited and etched to form an anti-reflection coating (ARC) layer on top of the Ge mesa for optical illumination.

### Measurement

The electrical and optical setups shown in Extended Data Fig. 1 are used for the counting and time-correlated counting in Fig. 3, and the 3D scanning and imaging in Fig. 4. All experiments are done in a standard laser lab environment at room temperature. A power supply provides a DC voltage to the DC input port of a bias-Tee (Anritsu K251). A pulse function generator (Keysight 81150A) provides a RF voltage pulse to the RF input port of the bias-Tee, and triggers the 1310 nm pulsed laser (Picoquant LDH with Taiko PDL driver) with a programmable time delay. The output of the bias-Tee is connected to a custom-made RF probe in which alternating-current (AC) bypass and 50 Ω termination are implemented. A low-noise amplifier (Mini-Circuits ZFL-1000LN+) is used to amplify the voltage signal to be fed to the counter (Picoquant Multiharp 150) or the oscilloscope (Keysight EXR254A). The output of the pulsed laser is 1 % split for optical power monitoring and 99 % attenuated by a variable optical attenuator (Thorlabs V1000PA). The output of the variable optical attenuator is 10 % split for optical power monitoring and 90% collimated by a collimator. The collimated laser pulses are injected into a custom-made confocal microscope, which consists of a pellicle beam splitter and an objective lens to respectively reflect and focus the laser pulses toward the device, and another pellicle beam splitter and two tube lenses to visually monitor the device with two cameras, one visible and another SWIR. When performing the counting using the counter (i.e., for count rate related measurements), the event pulse edge and the gate pulse edge are used to trigger and sync the counter, respectively. The counter integration time is typically set to 100 s. When performing the counting using the oscilloscope (i.e., for jitter related measurements), the rising edge of the event pulse is used to trigger the oscilloscope data capturing frame. Histograms are generated using the rising edge of the event pulse in reference to the gate pulse over 60 K events. When performing the time-correlated counting using the counter (i.e., for after-pulsing related measurements), the output of the low-noise amplifier is divided into two channels with equal power by a RF power divider (custom-made), and a small time delay due to a longer RF cable is introduced to one of the two channels. The outputs of the more delayed channel and the other less delayed channel are respectively coupled to the sync and trigger input ports of the counter to enable the time-correlated counting of interarrival time. The counter integration time is typically set to 200 s.

When performing the 3D scanning and imaging using the counter, the variable optical attenuator is replaced by a coaxial transceiver consisting of a circulator, a collimator, and x-y galvo-mirrors (Thorlabs QS30XY-AG). Note that a 1310 nm CW laser and a 635 nm CW laser are coupled to the same optical path of the pulsed laser, and respectively used to align the focused laser pulses with the device under the confocal microscope and determine the initial angular direction of steering the mirrors. The laser pulse width and peak power are estimated to be 30 ps and 5 mW, respectively. The mirrors are angular-direction-controlled between x-y angles equal to (-6 °, 4 °) to (9 °, -5 °) with 0.1 ° resolution in both directions in a feedback loop such that the laser pulses are fired according to the angular directions following the steering commands. At every steering command, the acquired histogram from the counter is analyzed and the temporal location of its peak is identified and translated into the corresponding x, y, z positions in the PCL dataset. $T_g$ and $T_r$ are set to be equal to 5 ns and 250 ns, respectively, with sufficient integration time for multi-shot LiDAR operation to overcome the high optical loss associated with the whole setup. Since $T_g$ is only 5 ns corresponding to the maximum TOF range limited to 75 cm, a space that spans roughly 0.5 $m^3$ is prepared and filled with a series of 3D objects for photo shoot, where the center of the space is located away from the mirrors by roughly 1.15 m.

Finally, the repeatability of DCR and TCR measurements done over 10 consecutive experiments using the counter is determined to be < 0.1 % error (standard deviation divided by mean) for both die-level testing and wafer-level testing; the accuracy of DCR and TCR measurements done at different times is estimated to be ~ 10 % error for die-level testing and < 10 % for wafer-level testing, possibly relating to the fluctuation of ambient temperatures.

## Calculation

By fitting $n_{AP}(i)$ with an exponential probability density function $\gamma e^{-t/\tau}$, $\tau$ in Fig. 3 (h) can be calculated by

$$\tau = \frac{T_r}{\ln(\frac{n_{AP}(1)}{n_{AP}(2)})} ,$$

and APPs shown in Fig. 3 (i) can be calculated by

$$APP(i) = \frac{n_{AP}(i)}{1+\sum_{i=1}^{\infty} n_{TG}(i)} \text{ and } APP^t = \frac{n_{AP}}{\left(1+\sum_{i=1}^{\infty} n_{TG}(i)\right)\cdot \frac{T_r}{T_g}} ,$$

where

$$n_{AP}(i) = \int_{iT_r}^{iT_r+T_g} \gamma e^{-t/\tau} dt \text{ and } n_{AP} = \int_0^{\infty} \gamma e^{-t/\tau} dt .$$

Here, the interpretation of $APP(i)$ is, under gated-mode operation, the probability of finding an after-pulsing event during the $i$th gate time when a dark count occurs during the reference gate time. And, the interpretation of $APP^t$ is, under free-running-mode operation, the total probability of finding an after-pulsing event when a dark count occurs.

**Acknowledgements**

We thank Dr. C. H. Shen and Mr. D. Y. Lai on the discussion over fabrication process and metrology.

**Author contributions**

N. N. conceived the design and supervised the project. Y.-C. Lu contributed to device simulation, layout synthesis, and wafer-level data analysis. Y.-H. L. contributed to layout synthesis, process integration, and wafer-level data analysis. P.-W. C. and Y.-C. Lai contributed to process integration and device fabrication. Y.-R. L. contributed to process integration and wafer-level data analysis. C.-C. L. contributed to wafer-level data collection and analysis. T. S. contributed to die-level data collection and analysis, and 3D PCL image capture. C.-H. C. contributed to die-level data collection and analysis. S.-L. C. oversaw the project.

**Competing Interests**

All authors are shareholders of Artilux Inc., a start-up company that makes SWIR 2D/3D sensors, imagers, and, photonic integrated circuits.

**Additional information**

**Correspondence and requests for materials** should be addressed to Neil Na.

**Data availability**

The measurement method and raw data supporting this work are available in the main text and extended data. Further additional information is available from the corresponding author upon request.

**Code availability**

The programs and codes supporting this work are used and written only for extracting and fitting the raw data from the measurement equipment, and are available from the corresponding author upon request.

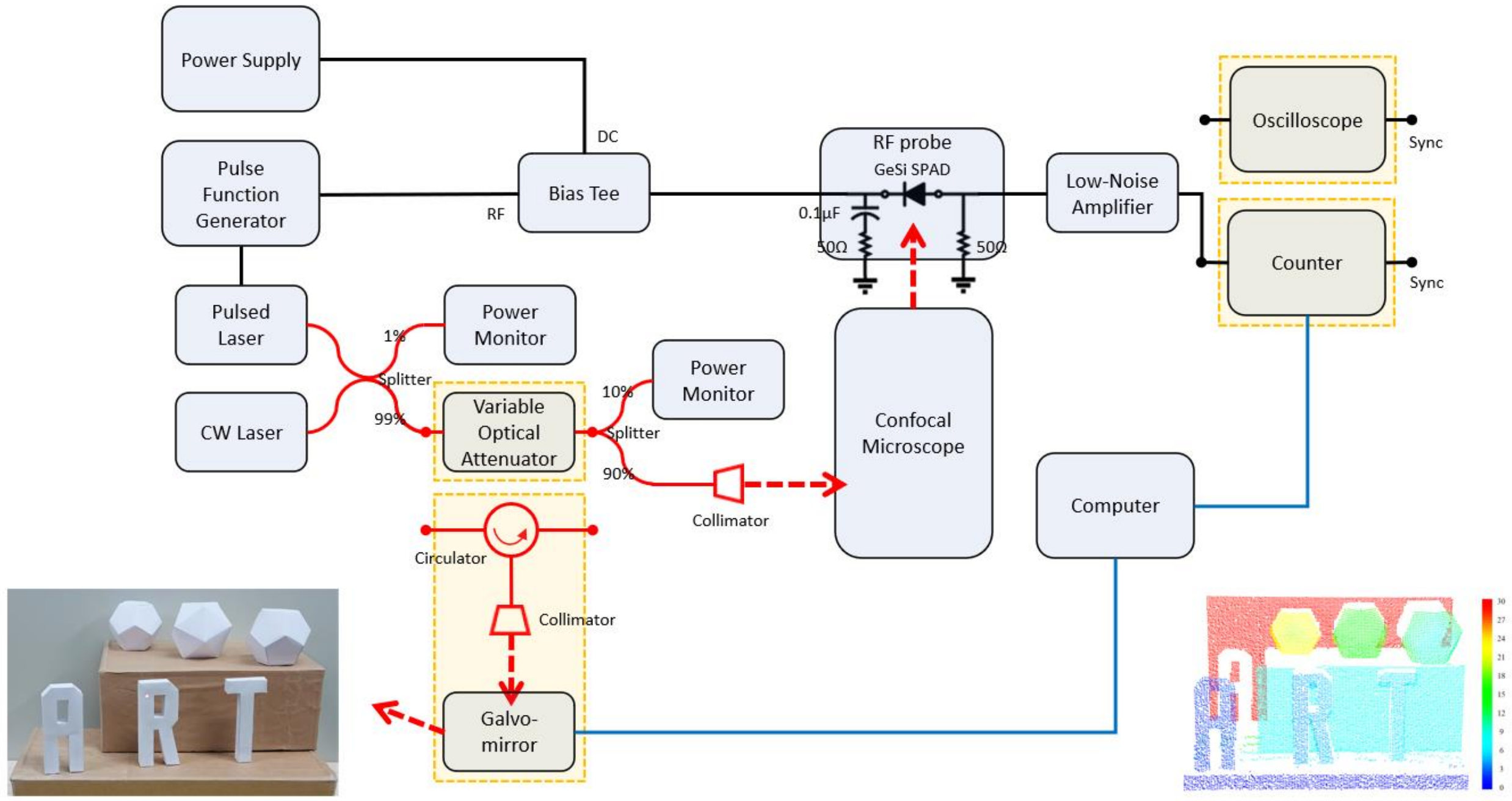

**Extended Data Fig. 1 | The schematic plot of the electrical and optical setups used in producing data shown in Fig. 3 and 4.**